# BLAST: Bridging Length/time scales via Atomistic Simulation Toolkit

Henry Chan[1,*], Badri Narayanan[1,2], Mathew Cherukara[1], Troy D. Loeffler[1], Michael G. Sternberg[1], Anthony Avarca[1], and Subramanian K. R. S. Sankaranarayanan[1,3,*]

[1] Center for Nanoscale Materials, Argonne National Laboratory, Argonne IL, USA
[2] Department of Mechanical Engineering, University of Louisville, Louisville KY, USA
[3] Department of Mechanical and Industrial Engineering, University of Illinois at Chicago, Chicago IL, USA

* Corresponding authors: Henry Chan (hchan@anl.gov) or Subramanian K. R. S. Sankaranarayanan (skrssank@anl.gov, skrssank@uic.edu)

## ABSTRACT

*The ever-increasing power of supercomputers coupled with highly scalable simulation codes have made molecular dynamics an indispensable tool in applications ranging from predictive modeling of materials to computational design and discovery of new materials for a broad range of applications. Multi-fidelity scale bridging between the various flavors of molecular dynamics i.e. ab-initio, classical and coarse-grained models has remained a long-standing challenge. Here, we introduce our framework BLAST (Bridging Length/time scales via Atomistic Simulation Toolkit) that leverages machine learning principles to address this challenge. BLAST is a multi-fidelity scale bridging framework that provide users with the capabilities to train and develop their own classical atomistic and coarse-grained interatomic potentials (force fields) for molecular simulations. BLAST is designed to address several long-standing problems in the molecular simulations community, such as unintended misuse of existing force fields due to knowledge gap between developers and users, bottlenecks in traditional force field development approaches, and other issues relating to the accuracy, efficiency, and transferability of force fields. Here, we discuss several important aspects in force field development and highlight features in BLAST that enable its functionalities and ease of use.*

# INTRODUCTION

Advances in high performance computing power, along with algorithmic improvements, have dramatically improved understanding of nanoscale materials phenomena across a broad range of applications from catalysis to energy storage to quantum information [1-5]. Despite significant improvements in materials modeling, our efforts to accurately capture atomic-scale dynamics, especially in low dimensional systems such as clusters and interfaces are still in their infancy. The interplay between often subtle dynamical processes at the nanoscale, including chemical reactions, transport of atoms and ions, defect chemistry, and solvation dynamics influences macroscopic observations and material properties [6-12]. A clear understanding of such dynamical in nanoscale clusters and across interfaces (e.g. solid-liquid) is crucial.

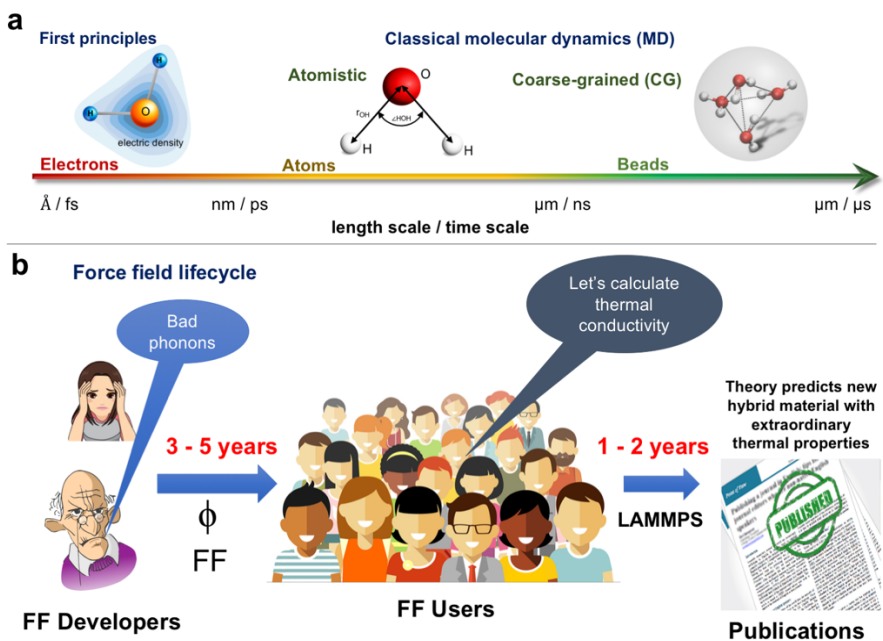

Figure 1. a) Length and time scale of molecular simulations. There is a balance between the accuracy and computational efficiency of different modeling techniques. b) Typical lifecycle of a force field. There are few force field developers who spent years developing one force field and know about its limitations, but there are thousands of users blindly taking the force field for molecular simulations.

Molecular dynamics (MD) simulations represent a popular method to understand such dynamical phenomena. First-principles approaches such as *ab initio* molecular dynamics (AIMD) can accurately capture diverse bonding environments within a single formulation. The AIMD simulations are computationally expensive and therefore limited to smaller systems (few hundred atoms) and short timescales (picoseconds). The dense linear algebra involved in DFT calculations makes it difficult to scale AIMD to much larger length scales limiting its ability to capture dynamical phenomena at typical nanoscale interfaces. While classical MD can tackle this challenge, its predictive power is limited by the interatomic potential models employed to capture various atomistic/ionic interactions. These potentials have been notoriously hard to develop especially for widely different (and often rapidly changing) chemical environments (including ionic, covalent, dispersive, hydrogen bonding, and electrostatic forces) typically exhibited by low dimensional systems and nanoscale interfaces. As such development of these interatomic potentials has always been a major challenge and represents an important aspect of multi-fidelity scale bridging (Figure 1a) that is at the heart of materials design and discovery.

A typical life cycle of the force-field development and usage is shown schematically in Figure 1b. Historically, there are only a handful of research groups that focus on development of interatomic potential models. These development efforts often entail several human years of effort to generate a single robust and accurate potential model for any given system. These models are released to a broad user community from academia and industry, who deploy them in materials modeling software/tools to execute MD simulations and study materials phenomena of their choice. Consequently, there is a disconnect between the force-field developers and the end users – in many cases, this leads to inadvertent use of force-fields wherein a user tries to model scenarios where an the underlying potential model either performs poorly or is simply incapable of modeling the underlying physical phenomena. Users typically do not have the flexibility to adapt, test, and validate these pre-defined potential models to problems of their interest.

Scientifically, existing force-fields have several limitations some of which are as follows: the training set generated in most cases does not adequately represent the potential energy space. It is quite common to over-emphasize equilibrium or near-equilibrium configurations, while those far-from-equilibrium and transition states are often ignored. Therefore, the models often poorly describe atomic-scale dynamics in scenarios where non-equilibrium configurations are encountered. Local optimization, and least squares are typical methods of choice for fitting the independent parameters. Initial guesses (or human intuition) play an important role in their success. Moreover, these methods are prone to convergence issues and over-fitting, especially in the large parameter space required for complex force-field formulations. Another major challenge is the substantial lack of quantitative cross-validation using properties that were not used in the fitting procedure. The description of the objective is also important; training procedures mostly use single objective function, (*i.e.*, a weighted sum of errors in prediction of target properties) to describe the quality of a given parameter set. The weights are arbitrarily chosen, primarily driven by intuition or experience. The resulting parameter search is inefficient, especially in a high-dimensional parameter space that is typical of modern-day force-fields.

The emergence of powerful machine learning (ML) techniques and advances in high performance computing hold tremendous promise in resolving the above challenges as well as the significant disconnect between force-field developers and the end-users. Here, we present our automated framework (BLAST – Bridging Length/Time scales via Atomistic Simulation Toolkit) that allows users to create their own models by generating training data sets, optimizing potential functions, and cross-validating their model predictions. BLAST employs a data-driven and ML based force field development approach that significantly deviates from the traditional approaches in several key aspects: the nature of training data set, the use of ML algorithms for advanced sampling and novel optimization schemes for fitting (*e.g.*, genetic algorithms, multi-objective optimization, neural networks to name a few), the form of the force fields, and cross validation and iterative improvement of ML based potential models. The details pertaining to the BLAST workflow, its objectives, implementation, user interface as well as representative test cases are presented in this paper. Finally, we also provide the outlook and future perspectives for BLAST development.

**BLAST WORKFLOW**

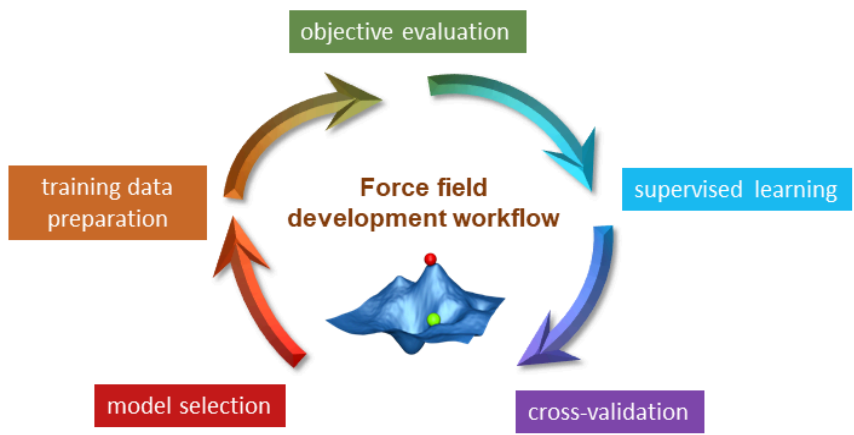

Figure 2. The major components in a force field development workflow: model selection, training data preparation, objective evaluation, supervised learning, and cross-validation. The major bottleneck of this workflow is the objective evaluation and the development time associated with data preparation and setup of the workflow.

A data-driven force field development workflow (Figure 2) consists of five main components: model selection, training data preparation, objective evaluation, supervised learning, and cross-validation. Below we briefly describe each of these components.

**1. Model selection**

A force field for molecular simulations is a model and its associated parameters. The model is ideally designed to capture all the physics necessary to describe the bonded and non-bonded interactions between atoms or coarse-grained beads. Traditionally, a model consists of analytical differentiable functions with up to hundreds of parameters, although recently there have been developments on more general/flexible force fields with up to millions of parameters that are based on complete basis functions or neural network with symmetry functions as the input layer. The choice of functional forms or neural network architectures defines the scope of the model parameter search space, and it also determines the efficiency and applicability of the force field, which imposes a ceiling limit on its accuracy.

## 2. Training data preparation

A data-driven approach in force field development relies on training data (*i.e.*, target data) collected from experiments or higher fidelity models. For instance, force fields for classical MD simulations are commonly trained against structures and energies calculated using first principle methods such as density functional theory. The quality of the training data establishes a baseline for the accuracy of a model, and it is important to sample data points that cover a wide distribution (*e.g.*, diverse structures that include different cluster size/shape, lattice type, chemical phases, etc.) to maximize the interpolate-ability of a force field. The quality of the developed force field generally improves with size of the training data but at the expense of longer training time, although recently active learning techniques coupled with on-the-fly data generation [*13*] have been shown to be an effective solution to this problem.

## 3. Objective evaluation

The quality of a force field candidate is quantified by the evaluation of an objective. For example, the objective in materials modeling is usually the sum of mean square errors between the predicted and target value of a list of material properties. These errors can be weighted or ranked and translated into a score that represents the performance of a force field candidate, which is optimized by a learning algorithm.

## 4. Supervised learning

At the heart of a data-driven learning approach is an effective strategy for navigating the model parameter search space. This process iteratively generates candidates and identify solutions based on the objective evaluation. Global search algorithms can explore wide regions of the search space, whereas local search algorithms are restricted in range but can effectively refine a solution.

## 5. Cross-validation

To alleviate the issue of overfitting, it is important to perform cross-validation using data or objectives that are outside of the training data. For example, in materials modeling, these can be structures excluded from the training or properties of the material predicted at conditions outside of the training.

## OBJECTIVES

The described force field development workflow is intensive in terms of both domain knowledge and development time. The objective of BLAST is to provide tools and software interfaces that simplify the workflow, such that even novice users can participate in the force field development process. Ultimately, the goal of BLAST is to leverage community-driven efforts to build a large collection of accurate yet computationally efficient force fields for molecular simulations across scales. Based on these objectives, we have established the following requirements for the design of BLAST:

## 1. Easy to use

BLAST should provide a user interface that is suitable for users with diverse computational skills, from novice users with no background in force field development to power users with domain knowledge. The interface should simplify labor-intensive processes in the workflow and address challenges relating to defining the scope of model parameter space, preparing training data, and designing the criterion for quality evaluation.

## 2. Easy to develop

In order to support a vast number of force field models and material properties, the functionality of BLAST should be easily extendable by developers. Therefore, the software architecture of BLAST should be modular, and if possible, adapt a standard data structure as its data storage system. The architecture should also be flexible enough to handle customization (*i.e.*, new custom models and material properties). The software should provide application interfaces (*i.e.*, wrappers) that enables integration with existing materials modeling software, particularly those with application programming or command line interfaces.

## 3. Massively parallelizable

The backend of BLAST should be tightly integrated with high performance or distributed computing resources, such that automations and parallelism can be used to significantly reduce the typical force field development time. The backend should implement common distribution protocols for handling both shared and distributed-memory systems, and support efficient parallel search and sampling algorithms as well as robust integration with existing highly parallelized energy evaluation codes such as LAMMPS, VASP, etc.

## WORKFLOW IMPLEMENTATION IN BLAST

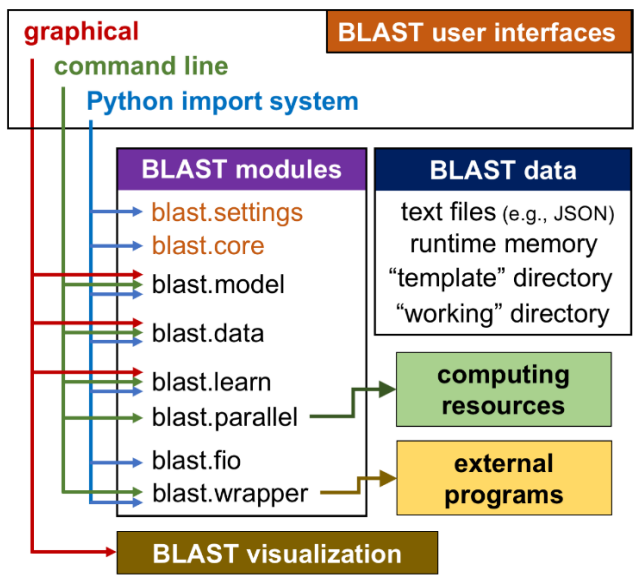

Figure 3. Software organization of BLAST. The software consists of Python modules which are accessible via user interfaces. BLAST implement three different user interfaces (*i.e.*, graphical, command line, Python import system) that are suitable for novice users as well as users with expertise in force field development. The web-based graphical interface also links to a visualization system that display statistical information using plots and graphs. BLAST uses text files (e.g., JSON) for internal data passing, and it also contains a template directory for storing input scripts. Within BLAST, there are core modules and a settings namespace that is shared among all BLAST modules. The model, data, and learn modules corresponds to the model selection, training data preparation, and supervised learning components of the force field development workflow. In addition to these main modules, the fio module handle all file input/output, the parallel module handle the running of BLAST on massively parallel and distributed computing resources, and the wrapper module provide application interfaces that connect to external command line programs such as MD simulators, density functional calculation software, etc.

The typical major bottleneck in force field development is the time associated with data preparation, setup of the workflow, and objective evaluation. Since program runtime excluding objective evaluation represents only a small fraction of the workflow, we mainly implemented the backend of BLAST in the Python programming language (*i.e.*, an interpreted language) for its flexibility as well as ease of development. To ensure data portability, BLAST mostly utilizes plain text files such as JSON for data storage, which can be easily serialized and streamed within and across the user interfaces and backend programs. Figure 3 illustrate the software design of BLAST, which consists of the following modules:

## blast.core

This module contains general reusable codes (*e.g.*, class definitions, file objects, command line functions, etc.) that are shared among all modules in BLAST. The inclusion of a core module promotes the writing of recurring programming units that helps reduce redundancy in the codebase as well as bugs associated with writing of new codes.

### blast.settings

This module establishes a namespace in runtime memory for settings that can be accessed by modules in BLAST through the Python import system. It provides mechanics to import settings from JSON files into its namespace during the initialization of BLAST.

### blast.model

This module maintains a database of force field model options. For each model, the database contains information about the name, bound, unit, and default range of the model parameters. BLAST implements existing functional forms (*e.g.*, in LAMMPS) and can be extended to support custom models.

### blast.data

This module establishes the standard for the organization of training data and maintains a list of existing material databases with the ability to pull data from them. This module is tightly integrated with the graphical interface, which provides users the ability to create their own data set and leverage visualization to display graphs for error checking the data set.

### blast.learn

This module contains various artificial intelligence and machine learning algorithms for the supervised learning process. BLAST implements various two-stage (global and local) optimization strategies. The use of different optimization strategies depends on available resources and goals of the user.

### blast.parallel

This module integrates BLAST with computing resources. The module provides a parallelized map() function for distributing workloads to workers across computing nodes and a reduce() function for cumulating the results. This MapReduce approach ensures excellent scalability and enables the objective function of many force field candidates to be evaluated in parallel. The MapReduce functions in BLAST support multiple distribution protocols including multi-processing (MP), message passing (MPI), and network-based messaging (ZeroMQ). Parallelization within a computing node is done using multi-process methodology, which safeguards against concurrency related issues. Parallelization across multiple computing nodes is implemented via a variation of the Broker pattern which consists of a broker that mediate the communications among independent workers.

### blast.fio

This is a support module for handling the input/output of different file formats.

**blast.wrapper**

This module interfaces BLAST with other external programs such as MD simulators, free energy calculators, density functional software, etc.

**"template" directory**

This is a directory within BLAST for storing commonly used input scripts, which ease the BLAST program setup by user or initialization of different modules. This directory is usually accessed by the wrapper module.

The development of BLAST is managed using the GIT version tracking software. To improve the clarity, we follow the conventional commits specification for commit messages. The software versioning of BLAST follows the major.minor.patch numbering system where major number corresponds to minor backward-incompatible (*i.e.*, breaking) changes, minor number corresponds to feature addition, and patch number corresponds to hotfixes. To support a continuous development process, the master branch is always production ready, and the implementation and testing of new features is done via branching in GIT.

**USER INTERFACES**

Most of the BLAST modules can be utilized using the standard Python module import mechanism. This is suitable for power users who are experience in coding and would like to build their own workflow using the building blocks in BLAST. For novice and advanced users, BLAST provides a command line interface complemented by a graphical interface. The command line interface is implemented by entry points based on the python __main__.py file. For example,

```
# Display a help page showing other entry points
    user@cli$ python -m blast

# Display a list of supported models and sub-arguments
    user@cli$ python -m blast.model

# Display a list of learning strategies and sub-arguments
    user@cli$ python -m blast.learn

# Display the instruction to run BLAST using different parallelism protocols
    user@cli$ python -m blast.parallel
```

A subset of the BLAST functionalities is exposed to a web-based graphical user interface (figure 4), which is suitable for novice users with limited force field development experience. However, the graphical interface offers features such as search space definition and training data preparation that can simplify the workflow, which can be useful even for advanced users. The graphical interface of BLAST consists of the following components:

### 1. Model Selection

This page allow user to select a force field functional form and define the parameter search space for a selected set of chemical elements or user defined CG bead types.

### 2. Training Data

This page allow user to import or specify a list of material properties and target values for the training process. This list of weighted or ranked material properties, *i.e.*, single objective or hierarchical objective, defines the quality evaluation of a parameter set.

### 3. Machine Learning

This page provides options to navigate the parameter search space via a one-stage or two-stage process. Users can select a global optimization strategy for a broad search and/or select a local optimization strategy for fine-tuning.

### 4. Job Submission

This page displays a list of submitted jobs and their status (e.g., progress, run time, requested computational resources, etc.). A user can start, cancel, monitor, or restart a job via this page.

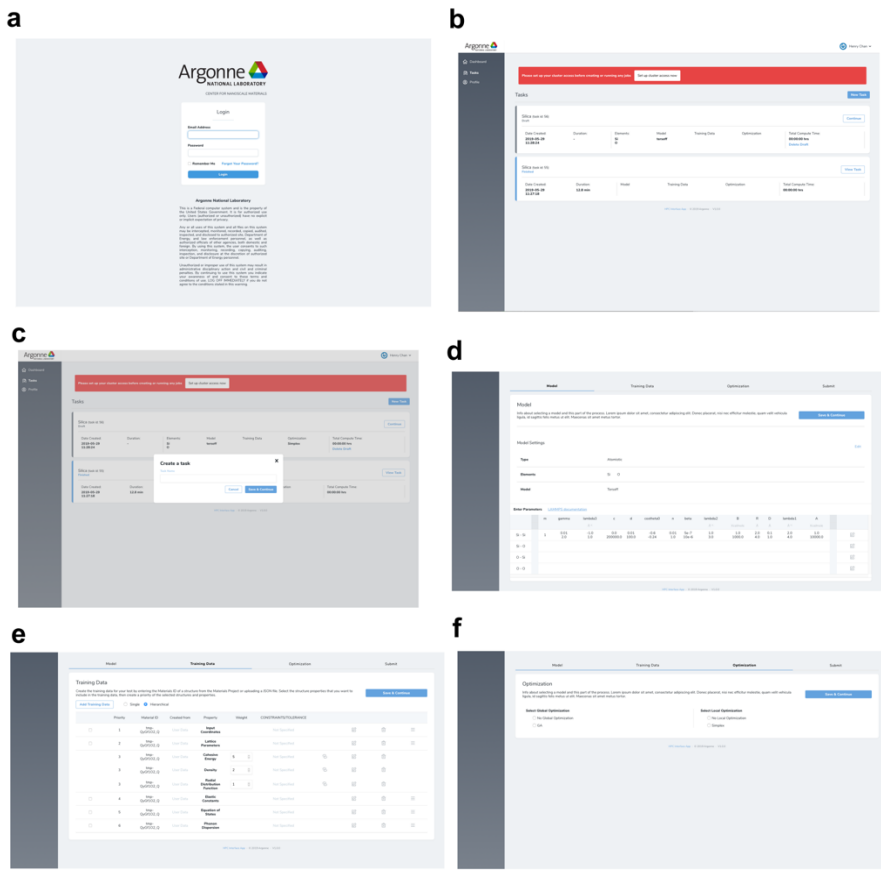

Figure 4. Example of the web-based graphical interface of BLAST. a) User login page. b) Dash-board style page for task management. c) Creation of new force field development task. d) Model selection page that allows users to easily define the scope of the model parameter search space. e) Training data page provides tools for users to create the training data as well as designing the objective evaluation (*i.e.*, weights, ranking, tolerance, etc). f) The optimization page provides options for different global and local optimization algorithms suitable and defines the runtime parameters of the supervised learning process.

# TEST CASES

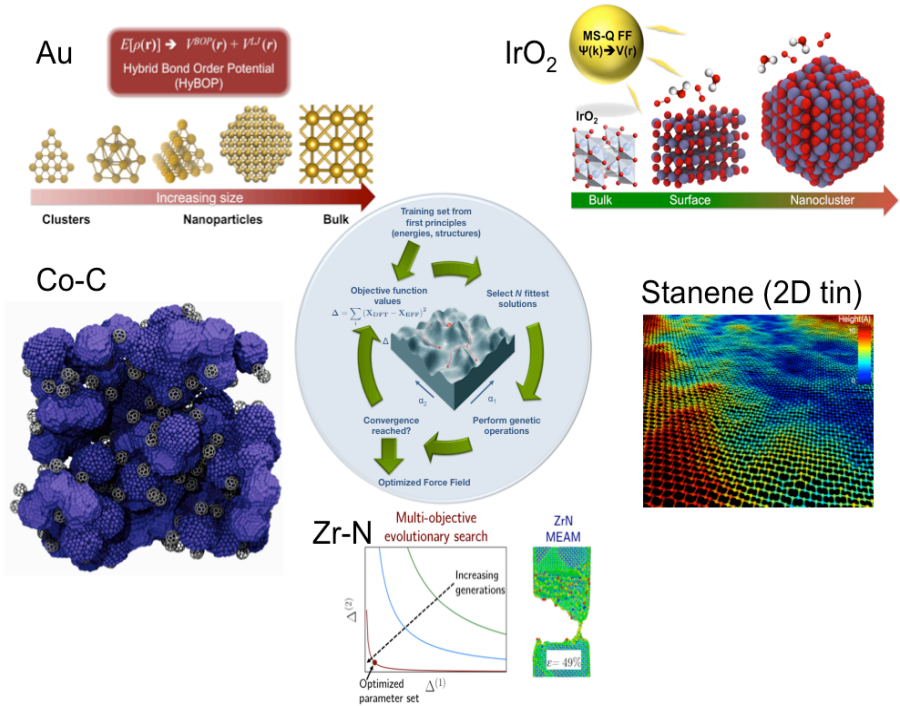

Figure 5. BLAST has had tremendous success in developing force fields for a wide range of materials from metals [14], oxides [15], nitrides [16], heterostructures [17], and even 2D materials [18-20]. Furthermore, BLAST has also been deployed to develop models for water [21-23] and other soft-matter systems.

BLAST has been deployed to exhaustively yet efficiently (low computational cost) scan the high-dimensional parameter landscape and obtain a set of parameters that best describe the properties chosen in the training data set. We have had extensive success in utilizing BLAST to develop accurate, transferable, and efficient interatomic potentials for a wide range of materials [24] (Figure 5), which includes metal nanoclusters [14], oxides [15], nitrides [16], and 2D materials [18-20] (stanene, silicene, and TMDCs such as $WSe_2$) to name a few representative systems. Our recently developed classical atomistic potentials accurately capture (a) structure, energetics, and dynamics across multiple length scales, (b) thermal properties, (c) interfacial reactions and synthesis pathways, as well as (d) mechanical properties and atomistic response to environment (temperature, pressure etc.). Furthermore, BLAST has also been used to develop a coarse-grained model to describe complex system such as water [21, 23] (Figure 6). The training set was chosen such that it provides ample representation of various regions of the potential energy landscape as well as dynamical properties. Specifically, we employed energies of thousand different configurations of ice obtained from the state-of-art all atom force field TIP4P. Since TIP4P fails to accurately capture dynamical properties, *e.g.*, melting point, we employed experimental data for such properties. We used experimentally measured (a) melting point, (b) density-temperature plots for ice and water, (c) dependence of specific heat of water on temperature, and (d) radial and angular distribution functions of water at ambient temperature. This is the first time wherein dynamical properties are explicitly fit in a rigorous fashion. BLAST allowed development of a coarse-grained model for water that accurately captures structure, dynamics, melting point, density and specific heat anomalies in water, and is quantitatively in exceptional agreement with experiments. In terms of accuracy in predicted properties, BLAST trained ML-BOP outperforms all the existing water models in its computational efficiency, predictive power, and ability to capture thermodynamic anomalies. Due to its coarse-grained description, we can employ timestep of ~10 fs for the MD simulations, which provides 3 times speed-up in terms of size (as compared to all-atom models; 3 atoms of water molecule is taken as 1 bead). Additionally, since the Tersoff formalism is a nearest neighbor potential, it is massively scalable. Such highly efficient formalism allowed us to model hitherto unexplored mesoscopic phenomena such as grain coarsening in ice under highly supercooled conditions.

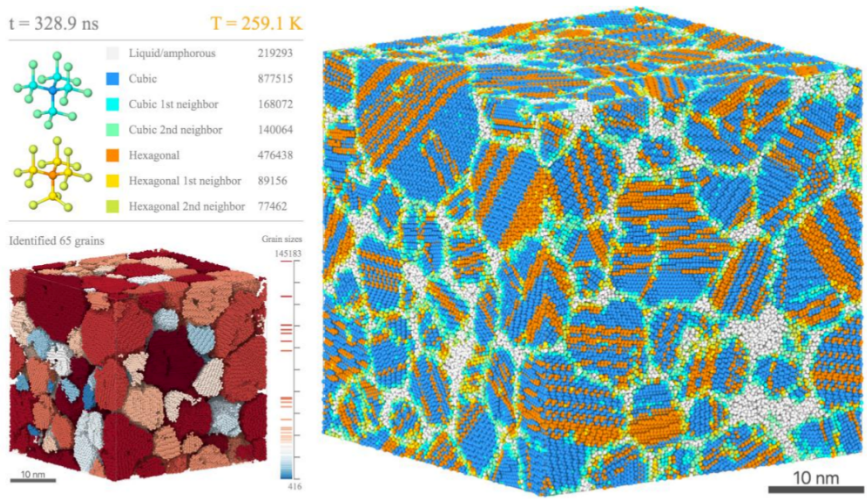

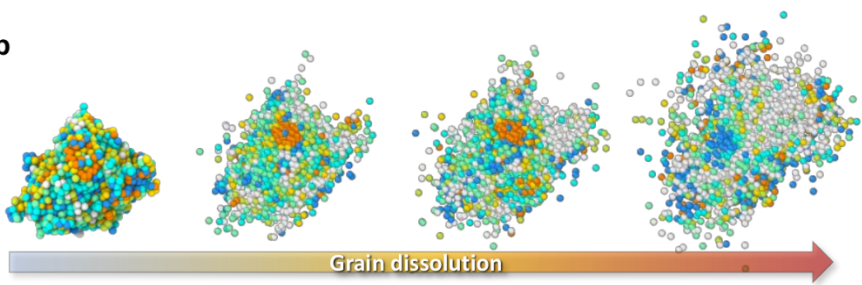

Figure 6. BLAST also can perform multi-fidelity scale bridging to train and validate coarse-grained models. An example of a water model [21] developed using BLAST is depicted above. (a) Snapshot from MD simulation of the annealing of polycrystalline ice. On the right, the water beads are colored by local structures (hexagonal, cubic, amorphous, etc.). On the bottom left, individual grains within the sample are identified using a clustering-based technique [25] and colored by their sizes. (b) The dissolution of a selected grain within the sample shown in (a).

# OUTLOOK AND FUTURE DEVELOPMENTS

BLAST is a powerful tool for materials modeling that leverages the recent advances in computing power. The computational efficiency of models developed using BLAST can accelerate materials discovery and design, which can be leveraged for advanced nanoscale materials characterization. BLAST addresses one of the most critical aspect of MD, *i.e.*, its predictive power which hinges strongly on the nature of the interatomic potential used to describe the atomistic interactions in the system. BLAST and the examples presented above highlight the advantages of using ML and data driven approaches for training new models. There is however still a lot of room for improvements. Future versions of BLAST will focus on addressing some of these issues as discussed below.

MD simulations that employ pre-defined functional forms remain popular owing to their computational efficiency but are limited in their ability to capture the underlying physics being modeled. BLAST interfaces with LAMMPS and allows MD users to choose from the list of pre-defined functions available within LAMMPS. There is always be a ceiling limit imposed by using pre-defined functional forms despite all the advances in both the extent of training data and the optimization algorithms. One needs flexibility in the functional form to overcome this challenge. BLAST future versions will implement neural network models to provide more flexibility to users interested in modeling complex reactive systems that involve multiple bonding characteristics (e.g. metallic, covalent and ionic).

All ML frameworks, including BLAST, require carefully curation and sampling of training data. In the context of materials models, one should ensure that the training data used is sufficiently diverse *(e.g.,* training set that consists of configurations spanning broad range of energies from near equilibrium to highly non-equilibrium). A developer must ensure ample representation of the different parts of the potential energy surface in the training data for the model to be robust. BLAST allows users to directly import first principles training data directly from Materials Project and thereby take full advantage of the vast amount of first-principles data available at our disposal. Future versions of BLAST will allow direct interface with electronic structure codes (e.g. VASP) to allow users to generate training data on-the-fly.

BLAST allows users to develop their own coarse-grained models using training data sampled from atomistic simulations and thus addresses an important time-scale challenge affecting classical MD simulations. All-atom simulations typically employ 0.1-1 femtosecond timesteps and as such are suited for modeling phenomena in the nanosecond timescales. Even with exascale computing, one would majorly gain in terms of spatial scales that can be accessed. The clockspeeds and bandwidths, however, are not going to change significantly and timescale challenges will remain. Coarse-grained models that typically allow 10-50 femtosecond timesteps can address this timescale challenge *albeit* typically at the cost of accuracy. BLAST enable users to overcome the timescale challenges without sacrificing accuracy. For example, the water model [*21*] developed using BLAST captures the thermodynamic properties and anomalies of water at fraction of the computational cost of its atomistic counterpart. BLAST enables development of accurate coarse-grained models that can model dynamical phenomena in the mesoscopic regime.

One of the major features in BLAST is that it allows users to include temperature dependent properties in its objective function. Traditionally, force field developers define objective functions in terms of static properties. Such models however lack predictive power to capture dynamical and transport properties. BLAST allows temperature dependent properties derived from on-the-fly MD to be included as part of the training procedure. BLAST users can directly train MD potentials that capture transport properties and other dynamical or temperature dependent properties of interest. Going forward, BLAST will include capability to include forces as part of the training data set.

BLAST upgrades will allow users to cross validate and iteratively improve their force-fields *via* active learning. BLAST will allow users to estimate error propagation from high-fidelity first-principles scale to coarse-grained mesoscopic scale. With the increasing use of higher-level theory (CCSD or QMC) to generate the training data and reduce the error necessitates quantification of uncertainties in the predictions at various scales. In this regard, Functional Uncertainty Quantification method (FunUQ) to assess model form errors, such as interatomic potentials for MD simulations represents an important future direction. Cross-validation, sensitivity analysis, and uncertainty quantification will be critical additions to BLAST, which enable users to test the robustness of their developed models.

BLAST uses ML global optimization algorithms and multi-stage optimization strategies that clearly represent a major advance over local optimization procedures that are commonly employed. Evolutionary optimization procedures such as genetic algorithms remain attractive, but one can easily envision the use of other emerging AI techniques such as Monte Carlo Tree Search (MCTS) to derive globally optimal solutions. BLAST upgrades will include the use of decision trees to navigate more effectively through the potential energy landscape. Finally, BLAST allows users to define objective functions that go beyond the simple sum of square difference. In addition to single objective functions that pose problems associated with selection of the appropriate weights, BLAST allows users to employ multi-objective optimization such as pareto optimization and HOGA (hierarchical objective genetic algorithm) to find optimal solutions that represent a compromise between the various desired objectives. Users will however need to exercise caution in defining the objective appropriately, otherwise, even the best global optimization strategy may not yield the best results.

## ACKNOWLEDGEMENTS

Use of the Center for Nanoscale Materials was supported by the U. S. Department of Energy, Office of Science, Office of Basic Energy Sciences, under Contract No. DE-AC02-06CH11357. This work utilized the Carbon cluster at the facility for the framework development.